\documentclass[acmsmall,screen]{acmart}

\AtBeginDocument{%
  }

\usepackage{enumerate}
\usepackage{enumitem}
\usepackage{multirow}
\usepackage{soul}
\usepackage{color, xcolor}
\usepackage{acronym}

\begin{document}

\acrodef{GNNs}{graph neural networks}
\acrodef{RNNs}{recurrent neural networks}
\acrodef{CNNs}{convolutional neural networks}
\acrodef{SSL}{self-supervised learning}
\acrodef{MCs}{Markov chains}
\acrodef{GRUs}{gated recurrent units}
\acrodef{FFTs}{Fast Fourier transforms}
\acrodef{DL}{Deep Learning}
\acrodef{MLP}{Multilayer Perceptron}

\title{Privacy-Preserving Sequential Recommendation with Collaborative Confusion}

\author{WEI WANG}
\email{202120421@mail.sdu.edu.cn}
\orcid{0000-0002-7080-3381}
\affiliation{%
  \institution{School of Information Science and Engineering, Shandong University}
  \streetaddress{72 Binhai Rd}
  \city{QingDao}
  \country{CHINA}
  \postcode{266237}
}

\author{Yujie Lin}
\email{yu.jie.lin@outlook.com}
\orcid{0000-0002-2146-0626}
\affiliation{%
  \institution{Zhejiang Lab}
  \streetaddress{Kechuang Avenue, Zhongtai Sub-District}
  \city{Hangzhou}
  \country{CHINA}
  \postcode{311121}
}

\author{Pengjie Ren}
\email{renpengjie@sdu.edu.cn}
\orcid{0000-0003-2964-6422}
\affiliation{%
  \institution{School of Computer Science and Technology, Shandong University}
  \streetaddress{72 Binhai Rd}
  \city{QingDao}
  \country{CHINA}
  \postcode{266237}
}

\author{Zhumin Chen}
\email{chenzhumin@sdu.edu.cn}
\orcid{0000-0003-4592-4074}
\affiliation{%
  \institution{School of Computer Science and Technology, Shandong University}
  \streetaddress{72 Binhai Rd}
  \city{QingDao}
  \country{CHINA}
  \postcode{266237}
}

\author{Tsunenori Mine}
\email{mine@ait.kyushu-u.ac.jp}
\orcid{0000-0002-7462-8074}
\affiliation{%
  \institution{Faculty of Information Science and Electrical Engineering, Kyushu University}
  \streetaddress{744 Motooka Nishi-ku}
  \city{Fukuoka}
  \country{JAPAN}
  \postcode{819-0395}
}

\author{Jianli Zhao}
\email{jlzhao@sdust.edu.cn}
\orcid{0000-0002-7291-9003}
\affiliation{%
  \institution{School of Computer Science and Engineering, Shandong University of Science and Technology}
  \streetaddress{579 Qianwangang Rd}
  \city{QingDao}
  \country{CHINA}
  \postcode{266590}
}

\author{Qiang Zhao}
\email{202120423@mail.sdu.edu.cn}
\orcid{0000-0001-6923-0080}
\affiliation{%
  \institution{School of Information Science and Engineering, Shandong University}
  \streetaddress{72 Binhai Rd}
  \city{QingDao}
  \country{CHINA}
  \postcode{266237}
}

\author{Moyan Zhang}
\email{zmy20001122@163.com}
\orcid{0000-0001-6130-1286}
\affiliation{%
  \institution{School of Information Science and Engineering, Shandong University}
  \streetaddress{72 Binhai Rd}
  \city{QingDao}
  \country{CHINA}
  \postcode{266237}
}

\author{Xianye Ben}
\email{benxianye@gmail.com}
\orcid{0000-0001-8083-3501}
\authornotemark[1]
\affiliation{%
  \institution{School of Information Science and Engineering, Shandong University}
  \streetaddress{72 Binhai Rd}
  \city{QingDao}
  \country{CHINA}
  \postcode{266237}
}

\author{Yujun Li}
\email{liyujun@sdu.edu.cn}
\orcid{0000-0003-4455-5991}
\authornotemark[1]
\affiliation{%
  \institution{School of Information Science and Engineering, Shandong University}
  \streetaddress{72 Binhai Rd}
  \city{QingDao}
  \country{CHINA}
  \postcode{266237}
}

\renewcommand{\shortauthors}{WANG et al.}

\begin{abstract}
Sequential recommendation has attracted a lot of attention from both academia and industry, however the privacy risks associated to gathering and transferring users' personal interaction data are often underestimated or ignored. Existing privacy-preserving studies are mainly applied to traditional collaborative filtering or matrix factorization rather than sequential recommendation. Moreover, these studies are mostly based on differential privacy or federated learning, which often leads to significant performance degradation, or has high requirements for communication. 

In this work, we address privacy-preserving from a different perspective. Unlike existing research, we capture collaborative signals of neighbor interaction sequences and directly inject indistinguishable items into the target sequence before the recommendation process begins, thereby increasing the perplexity of the target sequence. Even if the target interaction sequence is obtained by attackers, it is difficult to discern which ones are the actual user interaction records. To achieve this goal, we propose a \textbf{C}o\textbf{L}laborative-c\textbf{O}nfusion seq\textbf{U}ential recommen\textbf{D}er, namely CLOUD, which incorporates a collaborative confusion mechanism to edit the raw interaction sequences before conducting recommendation. Specifically, CLOUD first calculates the similarity between the target interaction sequence and other neighbor sequences to find similar sequences. Then, CLOUD considers the shared representation of the target sequence and similar sequences to determine the operation to be performed: keep, delete, or insert. We design a copy mechanism to make items from similar sequences have a higher probability to be inserted into the target sequence. Finally, the modified sequence is used to train the recommender and predict the next item. 

We conduct extensive experiments on three benchmark datasets. 
The experimental results show that CLOUD achieves a modification rate of 66\% for interaction sequences, and obtains over 99\% recommendation accuracy compared to the state-of-the-art sequential recommendation methods. 
This proves that CLOUD can effectively protect user privacy at minimal recommendation performance cost, which provides a new solution for privacy-preserving for sequential recommendation.

\end{abstract}


\maketitle

\section{Introduction}
As an effective tool for addressing information overload, recommendation systems \cite{1-wang, 2-zhao} have been widely used in network service platforms. Due to the dynamic changes of user interest characteristics, many works have been devoted to modeling the evolving patterns of historical user interactions. 
Among them, sequential recommendation has attracted more attention in recent years, as it enables the extraction of rich temporal information from user behavior records \cite{3-Eduardo,4-Ming}. 
The goal of sequential recommendation \cite{5-fang,6-wang,7-wang} is to capture the transfer relationship between items, so as to recommend the next item for users according to the ordered interaction records over a period of time. Early models are based on \acf{MCs} \cite{8-Florent,9-he,10-Rendle}. Later, deep learning becomes the mainstream method. Some of the most advanced architectures such as \acf{RNNs} \cite{11-hidasi,12-ma,13-Ren,14-wu}, \acf{CNNs} \cite{15-Tang,16-tuan,17-Yuan}, memory networks \cite{18-chen,19-Huang,20-wang}, \acf{GNNs} \cite{21-chang,22-wu,23-zhang} and transformers \cite{24-kang,25-sun,26-wu} have been combined for sequential recommendation tasks. More recently, \acf{SSL} has been introduced to sequential recommendation for extracting robust item correlations by semi-automatically exploiting raw item sequences \cite{27-ma,28-xia,29-xie,30-yu}.

At the same time, the privacy risks associated to gathering and transferring users’ personal interaction data  are attracting more attention. However, current research on sequential recommendation mostly focuses on improving accuracy \cite{24-kang, 25-sun, 29-xie}, diversity \cite{68-cen, 21-chang}, robustness \cite{50-zhou, 40-lin} and so on,  while neglecting the privacy-preserving of user interaction sequences. 
In addition, existing privacy-preserving research in recommendation systems is mainly based on differential privacy \cite{31-abadi,32-dwork,33-Shin} or federated learning \cite{34-wu,35-zhang,36-wu}. Most differential privacy-based research adds noise to the feature representation or gradient, making it difficult for attackers to obtain user's relevant knowledge through the recommendation output, however which may bring significant performance degradation \cite{37-zhu}. Federated learning-based research deploys and trains the recommendation model on users' local devices, while transfers parameters between local devices and the server, which requires extra cost on communication and deployment \cite{38-li}. 
For instance, \citet{39-han} propose DeepRec,  which is one of the few works that aims to improve the privacy-preserving performance of sequential recommendation. DeepRec first trains a global model based on existing user purchase records, and then fine-tunes the private model based on the new interaction records on the user’s device, thereby reducing the risk of the recommendation server leaking user privacy. Essentially, DeepRec uses distributed training methods and also imposes high requirements on communication and deployment.

\begin{figure}[h]
  \centering
  \includegraphics[width=0.75\linewidth]{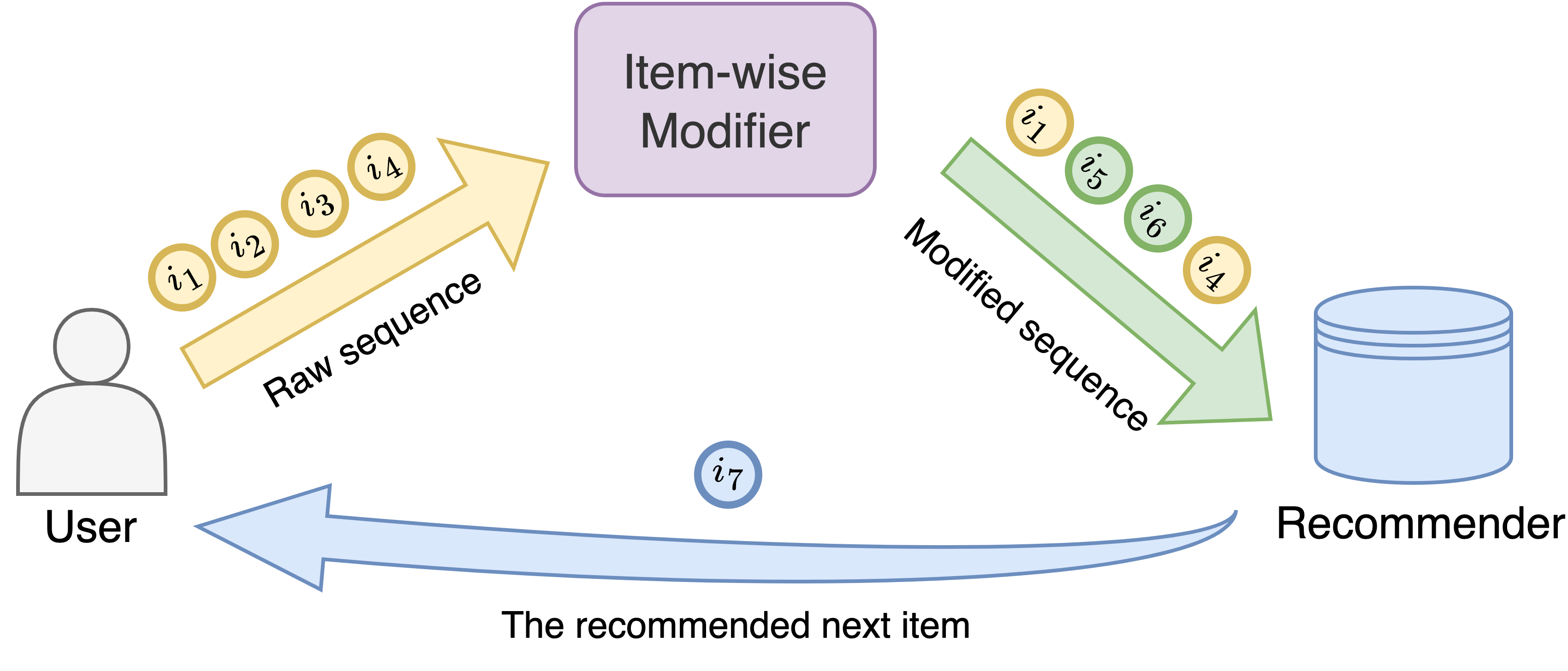}
  \caption{Schematic diagram of the recommendation process of CLOUD. The recommender only stores the modified interaction sequence, and even if it is attacked, the real interaction sequence will not be leaked.}
  \Description{The figure is used to help readers understand what is presented in the introduction.}
  \label{fig1}
\end{figure}

In this work, we attempt to address privacy-preserving in sequential recommendation from a different perspective.
Intuitively, we seek to modify the user's interaction data, delete some records, insert some fake records so that the recommendation server does not store the real user interaction sequence. 
As shown in Figure~\ref{fig1}, the user's real interaction sequence is modified by an item-wise modifier, and the recommender can only receive the modified interaction sequence. 
We delete some items and inject some indistinguishable items, thus significantly enhancing the perplexity of the interaction sequence. 
Even if the modified interaction sequence is obtained by attackers, it is difficult to distinguish which ones are the user's real interaction records, thereby reducing the risk of user privacy leakage.

To achieve this goal, we propose a \textbf{C}o\textbf{L}laborative-c\textbf{O}nfusion seq\textbf{U}ential recommen\textbf{D}er, CLOUD. CLOUD contains a collaborative confusion mechanism, which is used to increase
the perplexity of the interaction sequence. Specifically, CLOUD has an item-wise modifier and a recommender. 
The item-wise modifier takes several sequences with the highest similarity to the target sequence as collaborative signals, and refers to them for collaborative confusion. For each item in sequence, there are three types of operations: keep, delete, and insert. CLOUD first uses the shared feature representation of the target sequence and similar sequences to determine the operation type for each item. If the operation is `keep', the item-wise modifier keeps the item. If the operation is `delete', the item-wise modifier will delete the item directly. 
If the operation is `insert', the item-wise modifier will employ a reverse generator to determine which items should be inserted, and reversely inserts them to the front of the target item in turn. 
We design a copy mechanism here to make items from similar sequences have a higher probability of being inserted, which is conducive to the model to perform the insertion operation more actively. 
Two self-supervised tasks are designed to train the item-wise modifier, which randomly delete and insert items in the raw sequence, and ask the item-wise modifier to restore the modified sequence, so that the model can learn the ability to delete and insert items. 
Finally, the modified sequence and the raw sequence are used together to train the recommender by the masked item prediction task \cite{25-sun}. 
We combine the item-wise modifier and recommender via a shared encoder and train them using a joint loss. 
Extensive experiments on three real-world datasets show that CLOUD achieves a maximum modification rate of 66\% for the interaction sequence, while the recommendation accuracy is still comparable to the state-of-the-art sequential recommendation baselines. 
In the above process, the real data is only used as the ground-truth in the offline training phase, and the recommender predicts the next item based on the modified sequence in practical applications.

Compared with existing research, our work has the following advantages: 
\begin{enumerate}[label=(\roman*)]
    \item The item-wise modifier directly modifies the user interaction record, and the recommender does not rely on the real item sequence, which is more straightforward than adding noise in intermediate processes such as embeddings or gradients.
    \item Compared with federated learning-based methods, centralized model training has no additional requirements for communication and device.
    \item Benefiting from the powerful self-supervised signals provided by the \ac{SSL}-based modifier to the item representations, the modification operations do not decrease the recommendation accuracy, and even improve it in most experiments, allowing CLOUD to achieve superior performance while reducing the risk of user privacy leakage.
\end{enumerate}
To sum up, the main contributions of this work are as follows:
\begin{itemize}
    \item We propose a collaborative-confusion sequential recommendation model, a new solution for privacy-preserving for sequential recommendation which can refer to similar sequences to automatically modify the user's interaction sequence before recommendation.
	\item The item-wise modifier with the copy mechanism can make CLOUD perform the insertion operation more actively, thereby increasing the perplexity of interaction sequence.
    \item Extensive experiments on three real-world datasets demonstrate that CLOUD can effectively protect user privacy at minimal cost to recommendation performance.
\end{itemize}


\section{Related work}
In this section, we survey related works from two categories: sequential recommendation and privacy-preserving recommendation.

\subsection{Sequential recommendation}
In online services like e-commerce platforms, user interactions are typically fragmented and encompass valuable temporal information. Most of the traditional collaborative filtering recommendation models treat each interaction record equally, which is difficult to capture the temporal information behind user behavior \cite{39-han}. To address this limitation, sequential recommendation has emerged as a recent focal point of research. The core goal of sequential recommendation is to model the transition relationship between items and predict the next item based on the ordered user-item interaction records. 
Early sequential recommendation models are based on Markov chains.
\citet{10-Rendle} combine first-order \ac{MCs} and matrix factorization to capture the dynamic changes of user interactions. Based on this, \citet{9-he} adopt high-order \ac{MCs} to more preceding items and mine more complicated patterns. 
With the development of deep learning, advanced frameworks such as \acf{GRUs}, \acf{CNNs} and \acf{GNNs} have also been introduced for sequential recommendation. \citet{41-Hidasi} adopt \ac{GRUs} to transfer the features of input sequences and propose an adaptive sequential recommendation model GRU4Rec. \citet{42-li} enhance the GRU-based sequential recommendation with an attention mechanism to capture the user's current preferences more accurately. \citet{43-bogina} consider users’ dwell time on items. \citet{15-Tang} propose a CNN-based model to model sequence patterns in neighbors. \citet{18-chen} propose a memory mechanism \cite{44-weston}, designing a memory-enhanced neural network to utilize the user's history more effectively. \citet{22-wu} utilize the \ac{GNNs} to model more complex item transition patterns in user sequences.

The proposal of self-attention \cite{61-vaswani} provides powerful feature representation capabilities for embedding sequences. SASRec \cite{24-kang} is the early work that applies self-attention to sequential recommendation.
Later, inspired by bidirectional transformer \cite{69-BERT}, BERT4Rec models the transition relationship of items in both left-to-right and right-to-left directions \cite{25-sun}. Self-supervised learning \cite{45-liu} has also been recently introduced to sequential recommendation due to its effectiveness in extracting contextual features. Based on SASRec, \citet{29-xie} use item crop, item mask, and item reorder as data augmentation approaches to construct self-supervision signals. \citet{46-liu} suggest utilizing data augmentation methods to construct self-supervised signals in order to effectively leverage item correlations. \citet{47-Qiu} conduct contrastive self-supervised learning using dropout.

Moreover, some denoising sequential recommendation works are also relevant to our study because they focus on removing noisy information in the sequences. \citet{48-qin} design a next basket recommendation denoising generator based on contrastive learning to determine whether the items in a historical basket are related to the target item. \citet{49-tong} leverage sequential patterns as prior knowledge to guide the contrastive policy learning model for denoising and recommendation. Inspired by \acf{FFTs}, \citet{50-zhou} propose an all-MLP model with learnable filters for denoising sequential recommendation.

Recently, \citet{40-lin} propose a self-correcting sequential recommendation algorithm named STEAM, which is relatively similar to our work. STEAM is designed to improve recommendation performance by removing misclicked items from the sequence and inserting missed items. However, STEAM only considers a single target sequence during correction operations, which provides less information to make the model more incline to keep items. Therefore, STEAM only has a low modification ratio which limits its ability to privacy protection.

Different from our work, the works mentioned above only focus on improving the performance of recommendation and do not consider privacy-preserving. Although denoising can modify the item sequences, they can only delete items and cannot insert new items, so the modified sequences are still users' real interaction records. Although STEAM has realized the modification of the input sequence, its modification ratio is very low, which is far from enough to confuse users' interaction records. We will analyze this in the experimental section.

\subsection{Privacy-preserving recommendation}
Most recommender systems rely on users' real interaction records to make recommendations. In practical commercial applications, unprotected data transmission brings the risk of user privacy leakage. With the rapid development of recommender systems, privacy-preserving in recommendation has attracted more attention. 
Up to date, most existing studies are based on differential privacy or federated learning. 
Differential privacy is widely used to disturb statistical data and feature information in recommendation. \citet{51-Arnaud} first apply differential privacy to matrix factorization recommendation and evaluate the performance difference of input noise at different stages. \citet{52-liu} combine differential privacy with bayesian posterior sampling to provide well performance. \citet{53-wang} apply differential privacy to domain-based recommendation by calibrating Laplacian noise into the training process. 
Federated learning saves user data on the user's local device and trains the model in a distributed manner, thereby reducing the risk of leaking data from the server. \citet{54-ammad} propose FCF, a classical federated collaborative filtering approach. FCF maintains user embeddings locally, item embeddings at the server, and stores user ratings locally for computing gradients of embeddings. Subsequently, \citet{55-Chai} propose a federated learning matrix factorization algorithm, which uses homomorphic encryption to protect item embeddings, so as to avoid leaking private rating information from the gradient of item embeddings. \citet{56-qi} further combine differential privacy and federated learning to encrypt the model gradient uploaded to the server, and propose a news recommendation model with privacy-preserving.

In general, most differential privacy-based methods add noise to the embedding or gradient, which makes it difficult for the attacker to obtain the original input of the user from the recommended output. This method perturbs the intermediate parameters of the model, and sometimes leads to a significant drop in recommendation accuracy \cite{37-zhu}. The federated learning-based methods train the model on users' local devices, which theoretically have a higher level of privacy-preserving. However, in the actual deployment, distributed training puts forward higher requirements for communication and users' devices, and also limits the recommendation performance \cite{38-li}.

In addition, the above studies are mainly applied in recommendation based on collaborative filtering or matrix factorization.
Up to date, there is few research devoted to protecting user privacy in sequential recommendation.
\citet{39-han} propose DeepRec, which is explicitly used to enhance privacy in sequential recommendation. DeepRec is a semi-distributed method, the server first trains a global model based on the existing user interaction data, then sends the recommendation parameters to user devices, combines the users' local interaction records to train a personal model and make recommendations. Compared with our work, the main limitation of DeepRec is that the distributed training of personal models significantly increases the communication cost and time cost, and also limits the complexity and recommendation performance of the model.

To sum up, we propose a new collaborative-confusion sequential recommendation model named CLOUD to enhance privacy-preserving. Compared with the above studies, CLOUD only uses the modified sequence to predict the next item. Since the real item sequence will not be stored in the server, the risk of user privacy leakage is reduced. In addition, modifying the raw sequences helps the recommender make more accurate predictions, which makes CLOUD take into account both privacy-preserving and recommendation performance. We believe that this work provides a new solution for privacy-preserving in sequential recommendation.

\section{Methods}
In this section, we introduce CLOUD in detail.
\begin{figure}[h]
  \centering
  \includegraphics[width=1.03\linewidth]{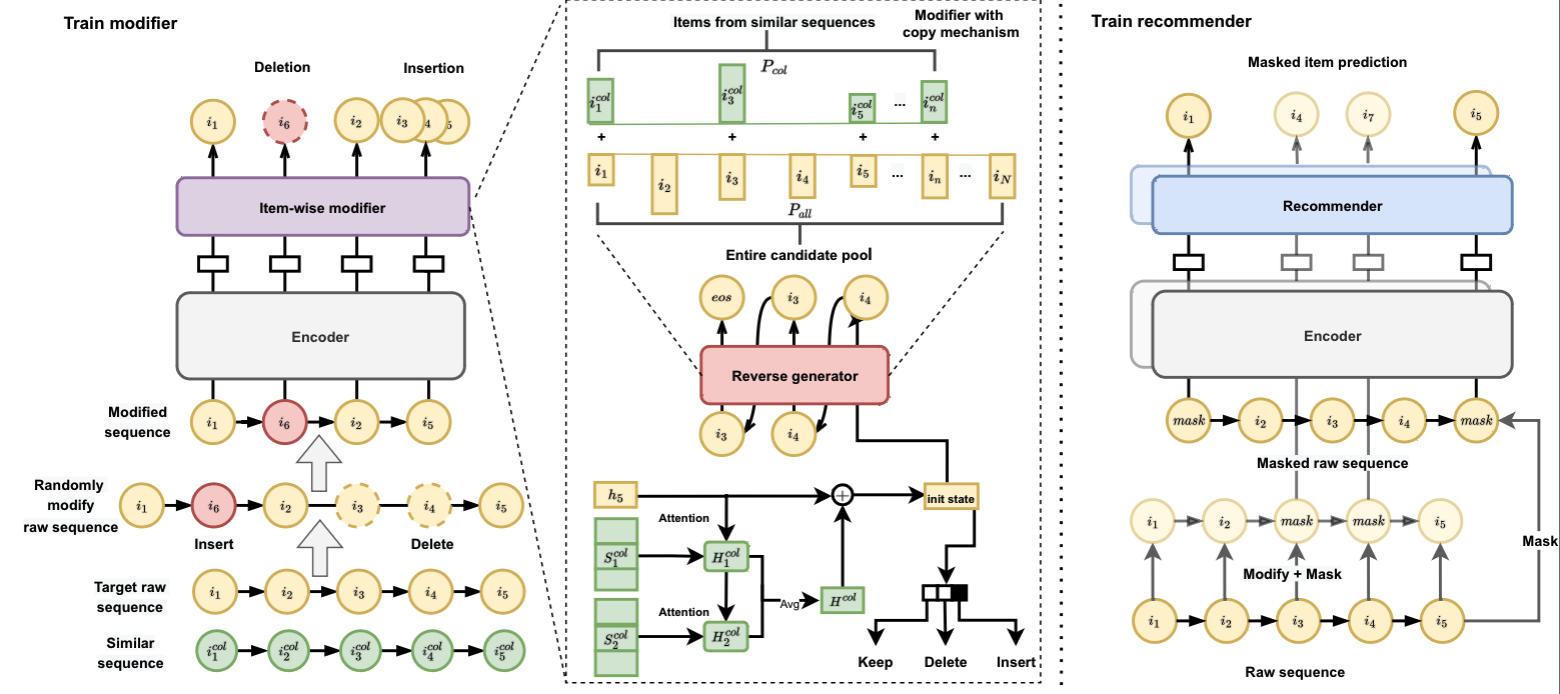}
  \caption{Overview of CLOUD. The encoder is used to encode the target raw sequence and similar sequence into hidden representation. When training the item-wise modifier, it is required to perform delete operation and insert operation to restore the randomly modified sequence. The target sequence and its modified version are used together to train the recommender. Finally, CLOUD jointly optimizes the loss functions of the item-wise modifier and recommender.}
  \Description{The figure is used to help readers understand CLOUD model.}
  \label{fig2}
\end{figure}

\subsection{Overview}
As shown in Figure \ref{fig2}, CLOUD has three main modules: encoder, item-wise modifier, and recommender. The encoder is used to encode the raw sequence and similar sequences into embeddings. We denote the set of items as $I$ and the target input sequence as $S=\left[i_{1}, i_{2}, \ldots, i_{|S|}\right]$, where $i_{t}\in I$ represents the item that user interacted with at the $t$-th time step and $|S|$ represents the length of the sequence. We calculate the Jaccard similarity between $S$ and other sequences, and define the $K$ sequences with the highest similarity as similar sequences $S^{c o l}=\left[i_{1}, i_{2}, \ldots, i_{\left|S^{c o l}\right|}\right]$. The item-wise modifier is used for collaborative confusion; it first predicts the actions to be performed at each position in the sequence: keep, delete, and insert. Then for the position where needs to insert items, a reverse generator is adopted to predict the item subsequence to be inserted. We design a copy mechanism here so that items $i^{col} \in S^{col}$ from the similar sequences have a higher probability of being inserted. The `replace' operation is not considered in this study, as this can be achieved by `delete' at the $t$-th position and `insert' at the next position. To train the item-wise modifier, we define a raw sequence $S^{r}$, then randomly modify $S^{r}$ and ask the item-wise modifier to restore it. The randomly modified sequence is denoted as $S^{m}$. At each position in $S^{r}$, the probability of `keep' operation is $p_k$, the probability of `delete' operation is $p_d$, the probability of `insert' operation is $p_i$, and $p_{k}+p_{d}+p_{i}=1$. Note that we do not delete the last item of $S^r$, as it will make the model confusion with the next item prediction. For the position in $S^m$ where items need to be inserted, we define a ground-truth sequence $S^{<i}=\left[i_{1}, i_{2}, \ldots, i_{\left|S^{<i}\right|-1}, e o s\right]$ used to indicate the items which should be inserted at the $i$-th position, where $\left[eos \right]$ is a special token representing the ending. Finally, the trained item-wise modifier modifies $S^r$ and outputs the modified sequence $S^c$, which is input to the recommender along with raw sequence $S^r$.

The task of the recommender is to predict the next item that the user is likely to interact with based on the input item sequence. Similar to the advanced sequential recommendation framework BERT4Rec \cite{25-sun}, the recommender is trained by the masked item prediction task. We randomly mask some items in the $S^r$ and $S^c$ with probability $p_m$, and ask the recommender to predict them, the masked items are replaced with a special token $\left[mask\right] \in I$. The masked $S^r$ and $S^c$ are denoted as $\widetilde{S^{r}}$ and $\widetilde{S^{c}}$, and the masked items are denoted as $\widetilde{I^{r}}$ and $\widetilde{I^{c}}$.

Finally, we train CLOUD by the joint loss function. The optimization goal of item-wise modifier is to maximize $P\left(S^{r} \mid S^{m}\right)=P\left(O \mid S^{m}\right) \times \prod_{i \in I^{i n s}} P\left(S^{<i} \mid S^{m}\right)$, where $O$ denotes the type of operation, $I^{ins}$ denotes the items whose ground-truth operations are `insert'. The optimization goal of recommender is to maximize $P\left(\widetilde{I^{r}} \mid \widetilde{S^{r}}\right)$ and $P\left(\widetilde{I^{c}} \mid \widetilde{S^{c}}\right)$. For testing, we add a $[mask]$ to the end of the modified sequence and ask CLOUD to predict the next item. The main mathematical notations used in this paper are summarized in Table \ref{tab1}. Next, we present the details of the three modules. 
\begin{table}[h]
	\centering
	\caption{Summary of main notations.}
	\label{tab1}
	\begin{tabular}{c|c}
		\toprule
		Notations & Definitions \\
		\midrule 
		$S$  &  target input sequence  \\
		$S^{col}$  &  similar sequence \\
        $I$ & the item set \\
		$i^{col}$  &  items from similar sequences \\
		$S^r$  &  raw sequence \\
		$S^m$  &  randomly modified sequence \\
        $S^c$  &  the sequence modified by the item-wise modifier \\
        $\widetilde{S^{r}}$ and $\widetilde{S^{c}}$ & the masked sequences \\
        $\widetilde{I^{r}}$ and $\widetilde{I^{c}}$ & the masked items \\
        $S^{<i}$ &  ground-truth sequence of insert operation \\
        $S_{1: n-1}^{<i_{t}}$ &  the currently generated inserted sequence \\
        $\left[eos\right]$ & the special token representing the ending \\

		$O$ &  the type of operation \\
		$p_k$  &  the probability of keep operation \\
        $p_d$  &  the probability of delete operation \\
        $p_i$  &  the probability of insert operation \\

        $E$ &  the item embedding matrix \\
        $e_t$ & the embedding vector of item $i_t$ \\
        $p_t$ & the position embedding at the $t$-th position \\
        $H_e$ & the hidden representation matrix of $S$ in encoder \\
        $H_{e^{col}}$ & the hidden representation matrix of $S^{col}$ in encoder \\
        $H_c$ & the hidden representation matrix of item-wise modifier \\
        $H_r$ & the hidden representation matrix of recommender \\

		\bottomrule 
	\end{tabular}
\end{table}

\subsection{Encoder}
The encoder is used to encode each position of the input sequence into a hidden representation. In CLOUD, the item-wise modifier and recommender share the encoder, that is, the output of the encoder is the input of the other two modules.

Specifically, encoder first defines the item embedding matrix $E \in \mathbb{R}^{|I| \times e}$ to project the representation of each item into a low-dimensional dense vector, where $e$ is the dimension of the embedding vector. For an item $i_t$ in the input sequence $S$, index the embedding matrix $E$ to obtain its embedding vector: $e_{t} \in \mathbb{R}^{e}$. We then inject position information into the model by adding a position embedding:
\begin{equation}
\label{eq1}
    h_{t}^{0}=e_{t}+p_{t},
\end{equation}
where $p_t$ represents the position embedding at the $t$-th position and $h_{t}^{0}$ represents the initial hidden representation of item $i_t$. In addition, we also refer to \cite{24-kang} to adopt dropout strategy to $h_{t}^{0}$. After stacking the initial representation vectors of all items in $S$, the initial representation matrix $H_{e}^{0} \in \mathbb{R}^{|S| \times e}$ can be obtained. We can also get the initial representation matrix $H_{e^{c o l}}^{0} \in \mathbb{R}^{\left|S^{c o l}\right| \times e}$ of similar sequence by the same way.

We follow BERT4Rec to update $H_e^0$ and $H_{e^{col}}^0$ by a bidirectional transformer \cite{25-sun,61-vaswani} with $L$ layers:
\begin{equation}
\label{eq2}
    H_{e}^{l}=\operatorname{Trm}_{b i}\left(H_{e}^{l-1}\right),
\end{equation}
\begin{equation}
\label{eq3}
    H_{e^{c o l}}^{l}=\operatorname{Trm}_{b i}\left(H_{e^{c o l}}^{l-1}\right),
\end{equation}
where $\operatorname{Trm}_{b i}$ denotes a bidirectional transformer block, $H_e^l$ and $H_{e^{col}}^l$ denote the representation matrix at the $l$-th layer. Finally, the encoder inputs the last layer of $H_e^L$ and $H_{e^{col}}^L$ into modifier. To simplify the notation, we omit the superscript $\text { (i.e., } \left.H_{e}, H_{e^{c o l}}\right)$ in the following sections.

\subsection{Item-wise modifier}
The item-wise modifier is the core of CLOUD and is used to collaborative confuse the encoded interaction sequence. Since STEAM \cite{40-lin} also realizes the modification of the input sequence and has the most superior performance, we first introduce the self-modifier of STEAM, and then introduce the collaborative-modifier of CLOUD for comparison.

\subsubsection{Self-modifier of STEAM}

\begin{figure}[h]
  \centering
  \includegraphics[width=0.6\linewidth]{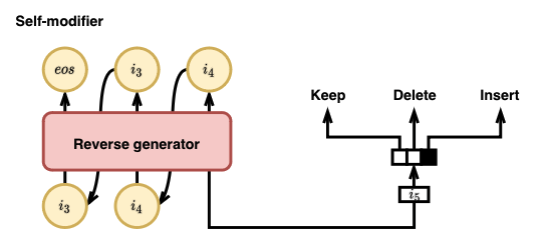}
  \caption{Self-modifier of STEAM. When inserting, the self-modifier uses a reverse generator to insert several items in reverse in front of the target position.}
  \Description{The figure is used to help readers understand the modifier of STEAM}
  \label{fig3}
\end{figure}

As shown in Figure \ref{fig3}, the self-modifier performs corrective operations at the item level. For each item in the raw sequence, it first selects one of the three operations: keep, delete, and insert. If the selected operation is `keep', the self-modifier skips the item. If the selected operation is `delete', the item is deleted directly. If the selected operation is `insert', the self-modifier uses a reverse generator to insert new items in front of the item. 

Given an item $i_t$ in the target sequence $S$ and its hidden representation $h_t \in {\mathbb{R}^e}$, STEAM follows Eq. \ref{eq4} to calculate the probability distribution of performing the three operations:
\begin{equation}
\label{eq4}
    P\left(\hat{o}_{t} \mid S\right)=\operatorname{softmax}\left(W h_{t}\right),
\end{equation}
where $\hat{o}_{t}$ denotes the predicted operation, $W \in \mathbb{R}^{3 \times e}$ is a projection matrix.

Assuming that an insertion operation needs to be performed before item $i_t$, the self-modifier applies a reverse generator to generate the item sequence to be inserted. The currently generated inserted sequence is denoted as $S_{1: n-1}^{<i_{t}}$. The self-modifier first indexes the embedding vector of each item in $S_{1: n-1}^{<i_{t}}$ from the item matrix $E$, then stacks the hidden representation $h_t$ of item $i_t$ and the embedding vector of each item in $S_{1: n-1}^{<i_{t}}$, while adding the position embedding:
\begin{equation}
\label{eq5}
    H_{c}^{0}=\left[\begin{array}{c}h_{t}+p_{1} \\e_{1}+p_{2} \\\ldots \\e_{n-1}+p_{n}\end{array}\right],
\end{equation}
where $H_c^0$ denotes the initial representation matrix of reverse generator. We also adopt dropout to $H_c^0$.

The reverse generator then updates $H_c^0$ by a unidirectional transformer to obtain its representation $H_c \in \mathbb{R}^{n \times e}$ at the final layer. Finally, STEAM follows Eq. \ref{eq6} to calculate the probability distribution of the next inserted item $i_n$:
\begin{equation}
\label{eq6}
    P\left(\hat{i}_{n} \mid S_{1: n-1}^{<i_{t}}\right)=\operatorname{softmax}\left(E h_{n}\right),
\end{equation}
where $E$ is the item embedding matrix, $h_n \in \mathbb{R}^e$ is the hidden representation of the last position of $H_c$. In particular, the first inserted item $i_1$ is generated based on $H_c^0=\left[h_t+p_1\right]$, so the probability is expressed as $P\left(\hat{i}_{1} \mid S\right)$.

The self-modifier only considers a single target sequence during correction operations, which provides less information to make STEAM more incline to keep items. Therefore, next we introduce a newly designed collaborative-modifier, which can significantly improve the modification ratio of the input sequence.

\subsubsection{Collaborative-modifier of CLOUD}

CLOUD introduces similar sequences and designs a new item-wise modifier to enhance the ability to modify the input sequence.

\begin{figure}[h]
  \centering
  \includegraphics[width=0.6\linewidth]{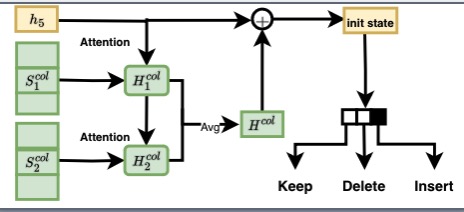}
  \caption{Schematic diagram of the probability distribution of three modification operations calculated by CLOUD.}
  \Description{The figure is used to help readers understand the item-wise modifier of CLOUD}
  \label{fig4}
\end{figure}

As shown in Figure \ref{fig4}, given a target sequence $S$ and its $K$ similar sequences $S^{col}$, CLOUD considers their information jointly to decide the action to be performed. We first follow Eq. \ref{eq7} and Eq. \ref{eq8} to obtain the shared representation $H_e^{\prime}$ of the target sequence and the similar sequences:
\begin{equation}
\label{eq7}
    a^{k}=\operatorname{softmax}\left(H_{e} \times\left(H_{e^{c o l}}^{k}\right)^{T}\right),
\end{equation}
\begin{equation}
\label{eq8}
    H_{e}^{\prime}=H_{e}+\left(\sum_{k=1}^{K} a^{k} \times H_{e^{c o l}}^{k}\right) / K,
\end{equation}
where $H_{e} \in \mathbb{R}^{|S| \times e}$ denotes the representation of target sequence $S$, $H_{e^{c o l}}^{k} \in \mathbb{R}^{\left|S^{c o l}\right| \times e}$ denotes the representation of $k$-th similar sequence, $a^k$ denotes the attention of $S$ to the $k$-th similar sequence calculated by Eq. \ref{eq7}.

Then, given item $i_t$ and its representation $h_t^{\prime} \in \mathbb{R}^e$ indexed from $H_e^{\prime}$, we follow Eq. \ref{eq9} to get the probability distribution $P\left(\hat{o}_{t} \mid S\right)$ of performing the three operations:
\begin{equation}
\label{eq9}
    P\left(\hat{o}_{t} \mid S\right)=\operatorname{softmax}\left(W h_{t}^{\prime}\right),
\end{equation}
where $\hat{o}_{t}$ denotes the predicted operation, $W \in \mathbb{R}^{3 \times e}$ is a projection matrix. When testing, the item-wise modifier performs the operation with the highest probability in $P\left(\hat{o}_{t} \mid S\right)$ for item $i_t$.

Assuming that item-wise modifier needs to perform an insertion before item $i_t$, CLOUD also applies a reverse generator to generate the items to be inserted. The currently generated inserted sequence is denoted as $S_{1: n-1}^{<i_{t}}$, and the initial hidden representation $H_c^0$ of reverse generator is obtained by Eq. \ref{eq5}. Then, we update $H_c^0$ with a unidirectional transformer, since generated items are not affected by later generated items:
\begin{equation}
\label{eq10}
    H_{c}^{l}=\operatorname{Trm}_{u n i}\left(H_{c}^{l-1}\right),
\end{equation}
where $\operatorname{Trm}_{u n i}$ denotes a unidirectional transformer block, $H_c^l \in \mathbb{R}^{n \times e}$ denotes the hidden representation matrix at $l$-th layer, we take the representation of the last layer and denote it simply as $H_c$.

\begin{figure}[h]
  \centering
  \includegraphics[width=0.6\linewidth]{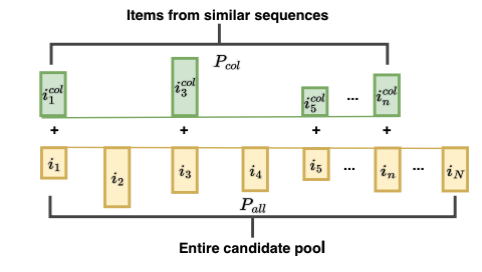}
  \caption{The copy mechanism of item-wise modifier. Items from similar sequences have a higher probability of being inserted overall.}
  \Description{The figure is used to help readers understand the item-wise modifier of CLOUD}
  \label{fig5}
\end{figure}

Next, the generator calculates the probability distribution for each item and predicts the next item to be inserted. As shown in Figure \ref{fig5}, we design a copy mechanism so that items from similar sequences have a higher probability of being inserted. Specifically, we apply an attention mechanism \cite{13-Ren} to match each similar sequence using the representation $h_n \in \mathbb{R}^e$ of the last position of $H_c$ to obtain an attention score:
\begin{equation}
\label{eq11}
    e_{k}^{n}=v_{c o}^{T} \tanh \left(W_{c o} h_{k}+U_{c o} h_{n}\right),
\end{equation}
where $W_{c o} \in \mathbb{R}^{e \times e}$, $U_{c o} \in \mathbb{R}^{e \times e}$ and $v_{c o} \in \mathbb{R}^{e \times 1}$ are transition matrices, $h_k \in \mathbb{R}^e$ denotes the average hidden representation of $k$-th similar sequence, $e_k^n$ denotes the attention score of $h_n$ to the $k$-th similar sequence. We normalize the attention scores and obtain the context feature vector $c^n$ from all similar sequences by weighted sum:
\begin{equation}
\label{eq12}
    a_{k}^{n}=\frac{\exp \left(e_{k}^{n}\right)}{\sum_{k=1}^{K} \exp \left(e_{k}^{n}\right)},
\end{equation}
\begin{equation}
\label{eq13}
    c^{n}=\sum_{k=1}^{K} a_{k}^{n} h_{k}.
\end{equation}

We use Softmax to convert $c^n$ and $h_n$ into a probability distribution to obtain the probability of inserting items from similar sequences $P\left(I^{c o l} \mid S^{c o l}, h_{n}\right)$ and the probability of inserting items from the full candidates $P\left(I^{all} \mid S^{c o l}, h_{n}\right)$:
\begin{equation}
\label{eq14}
    \left[P\left(I^{c o l} \mid S^{c o l}, h_{n}\right), P\left(I^{a l l} \mid S^{c o l}, h_{n}\right)\right]=\operatorname{softmax}\left(W_{c o}^{P} c^{n}+W_{a l l}^{P} h_{n}\right),
\end{equation}
where $W_{co}^P \in \mathbb{R}^{2 \times e}$ and $W_{all}^P \in \mathbb{R}^{2 \times e}$ are transition matrices, $I^{col}$ denotes the items from similar sequences and $I^{all}$ denotes all items in the dataset.

Finally, for each item in the candidate, we follow Eq. \ref{eq15} - Eq. \ref{eq17} to calculate the probability that it could be inserted:
\begin{equation}
\label{eq15}
    P_{c o l}\left(i_{j}\right)=\frac{\exp \left(e_{j} h_{n}\right) \times N}{\sum_{j}^{|I|} \exp \left(e_{j} h_{n}\right) \times N},
\end{equation}
\begin{equation}
\label{eq16}
    P_{a l l}\left(i_{j}\right)=\operatorname{softmax}\left(e_{j} h_{n}\right),
\end{equation}
\begin{equation}
\label{eq17}
    P\left(\hat{i}_{j} \mid S_{1: n-1}^{<i_{t}}\right)=P\left(I^{c o l} \mid S^{c o l}, h_{n}\right) \times P_{c o l}\left(i_{j}\right)+P\left(I^{a l l} \mid S^{c o l}, h_{n}\right) \times P_{a l l}\left(i_{j}\right),
\end{equation}
where $N$ indicates how many times item $i_j$ occurs in similar sequences, and $e_j \in \mathbb{R}^e$ represents the embedding vector of $i_j$ in the item embedding matrix $E$.

Since we have the ground-truth that which items need to be inserted when training, we can use the hidden representation of all positions of $H_c$ to calculate $P\left(\hat{i}_{j} \mid S_{1: n-1}^{<i_{t}}\right)$ for all $n$ at one time. When testing, the generator will start at the first position and predict the next inserted item in turn, until generating $\left[eos\right]$ or reaching the predetermined maximum insert length.

\subsection{Recommender}
Recommender is used to predict the masked items when training or predict the next item when testing. The recommender of CLOUD is replaceable, and this paper designs the recommender following the powerful framework BERT4Rec \cite{25-sun}.

Given the input sequence $S$ and its hidden representation matrix $H_e$ in the encoder, the recommender again updates $H_e$ by a bidirectional Transformer. We denote the initial hidden representation matrix of recommender as $H_r^0 \in \mathbb{R}^{|S| \times e}$, and $H_r^0=H_e$. $H_r^0$ is updated following Eq. \ref{eq18}:
\begin{equation}
\label{eq18}
    H_{r}^{l}=\operatorname{Trm}_{b i}\left(H_{r}^{l-1}\right),
\end{equation}
where $H_r^l$ denotes the representation matrix at $l$-th layer. We take the representation of the last layer and briefly denote it as $H_r$.

Suppose we mask the item $i_t$ at the $t$-th position in the sequence when training. The recommender follows Eq. \ref{eq19} to predict the probability distribution of the masked items:
\begin{equation}
\label{eq19}
    P\left(\hat{i}_{t} \mid S\right)=\operatorname{softmax}\left(E h_{t}\right),
\end{equation}
where $E$ denotes the shared item embedding matrix, $h_t \in \mathbb{R}^e$ denotes the hidden representation of the special token. The item $i_{t} \in \widetilde{I^{r}}\left(\widetilde{I^{c}}\right)$ and the sequence $S$ is $\widetilde{S^{r}}$ or $\widetilde{S^{c}}$. When testing, the masked item is added to the end of the modified sequence, $i_t$ is $i_{\left|S^{c}\right|+1}$ and the recommender calculates $P\left(i_{\left|S^{c}\right|+1} \mid S^{c}\right)$ to predict the next item.

\subsection{Loss function and model training}
We optimize CLOUD by training item-wise modifier and recommender.

To train the item-wise modifier, we first randomly perform deletion and insertion operations on a raw sequence $S^r$ to obtain the modified sequence $S^m$, and then ask the item-wise modifier to restore $S^m$ back to $S^r$. Specifically, for randomly inserted items, the modifier needs to accurately predict the deletion operation; For deleted items, the item-wise modifier is required not only to perform insert operation, but also to insert the correct items. We repeat this process for each raw sequence, enabling the model to be trained in a self-supervised manner without manual labeling. As shown in Eq. \ref{eq20}, the target loss function of the item-wise modifier is to minimize the negative log-likelihood of the probability $P\left(S^{r} \mid S^{m}\right)$:
\begin{equation}
\label{eq20}
    \begin{aligned}L_{m o d} & =-\log P\left(S^{r} \mid S^{m}\right) \\& =-\left(\log P\left(O \mid S^{m}\right)+\sum_{i \in I^{i n s}} \log P\left(S^{<i} \mid S^{m}\right)\right) \\& =-\left(\sum_{t=1}^{\left|S^{m}\right|} \log P\left(\hat{o}_{t}=o_{t} \mid S^{m}\right)+\sum_{i \in I^{i n s}} \sum_{n=1}^{\left|S^{<i}\right|} \sum_{j=1}^{|I|} \log P\left(\hat{i}_{j}=i_{n} \mid S_{1: n-1}^{<i}, S^{m}\right)\right)\end{aligned}.
\end{equation}

Then CLOUD adopts masked item prediction task to train the recommender. We randomly mask the raw sequence $S^r$ and the modified sequence, and ask the recommender to predict the masked items. The target loss function of the recommender is to minimize the negative log-likelihood of the probability $P\left(\widetilde{I^{r}} \mid \widetilde{S^{r}}\right)$ and $P\left(\widetilde{I^{c}} \mid \widetilde{S^{c}}\right)$:
\begin{equation}
\label{eq21}
    \begin{aligned}L_{r e c} & =-\left(\log P\left(\widetilde{I^{r}} \mid \widetilde{S^{r}}\right)+\log P\left(\widetilde{I^{c}} \mid \widetilde{S^{c}}\right)\right) \\& =-\left(\sum_{i \in \widetilde{I}^{r}} \log P\left(\hat{i}=i \mid \widetilde{S^{r}}\right)+\sum_{i \in \widetilde{I}^{c}} \log P\left(\hat{i}=i \mid \widetilde{S^{c}}\right)\right)\end{aligned}.
\end{equation}

Finally, we use the standard backpropagation algorithm to minimize the joint loss $L$ and optimize the parameters of CLOUD:
\begin{equation}
\label{eq22}
    L=L_{mod}+L_{rec}.
\end{equation}

\section{Experimental setup}
This section first presents four research questions that need to be answered experimentally, and then introduces the statistics and preprocessing of the datasets. Next, several state-of-the-art baselines are introduced. Finally, we present the evaluation metrics and parameter settings of experiments.

\subsection{Research questions}
We seek to answer the following research questions:

(RQ1) How much extent can CLOUD protect the privacy of user interaction sequence?

(RQ2) Does CLOUD have superior recommendation performance while protecting user privacy?

(RQ3) Will the recommendation performance of CLOUD be significantly affected when a large amount of noise is injected into the interaction sequence?

(RQ4) As the recommender in CLOUD is replaceable, do different recommenders significantly affect the performance of CLOUD?

\subsection{Datasets}
We conduct experiments on three datasets: Beauty, Yelp and Sports. Among them, Beauty and Sports are two types of product review datasets crawled from Amazon by \citet{62-McAuley}. Yelp is a business recommendation dataset published by Yelp.com. Since Yelp is very large, we only use the interaction records during 2019.

As with other works \cite{24-kang,40-lin,50-zhou}, we follow the common practice to preprocess the datasets. We first filter out users and items with less than 5 interaction records, and then sort the interaction records of each user in order of time to obtain the item sequences. For each item sequence, the last item is the test item, the second last item is the validation item, and the remaining items are used for training. In addition, we calculate the Jaccard similarity of every two sequences and select up to 10 sequences with similarity greater than 0.1 as co-sequences for each raw sequence. The statistics of the preprocessed datasets are shown in Table \ref{tab2}:

\begin{table}[h]\small%
	\centering
	\caption{Statistics of the datasets.}
	\label{tab2}
	\begin{tabular}{l|c|c|c}
		\toprule
		Datasets   &    Beauty & Yelp & Sports\\ 
		\midrule 
		Users & 22,362 & 22,844  & 35,597 \\
		Items & 12,101 & 16,552  & 18,357    \\  
		Records  & 194,682 & 236,999  & 294,483    \\
		Avg. length & 8.7  & 10.4 & 8.3\\
		Density & 0.07\%  & 0.06\%  & 0.05\%\\
		\bottomrule 
	\end{tabular}
\end{table}

\subsection{Baselines}
To verify the performance of CLOUD, we compare it with the following SOTA sequential recommendation baselines, which can be classified into three groups: (1) vanilla sequential recommendation models; (2) SSL-based sequential recommendation models, and (3) denoising sequential recommendation models.
\begin{itemize}
    \item Vanilla sequential recommendation models:
    \begin{itemize}
        \item GRU4Rec \cite{41-Hidasi} is an early work that adopts a gated recurrent network to predict the next item.
        \item SASRec \cite{24-kang} uses a unidirectional transformer module to model the transition relationship between items.
        \item BERT4Rec \cite{25-sun} introduces BERT into sequential recommendation, using a bidirectional transformer and trained by masked item prediction task.
        \item SRGNN \cite{22-wu} models item sequences using a graph neural network with an attention network.
    \end{itemize}

    \item SSL-based sequential recommendation models:
    \begin{itemize}
        \item CL4SRec \cite{29-xie} uses three self-supervised tasks based on item crop, item mask, and item reorder respectively to train a transformer-based sequential recommendation model.
        \item DuoRec \cite{47-Qiu} employs an augmentation method based on dropout and a novel sampling strategy to construct contrastive self-supervised signals.
        \item STEAM \cite{40-lin} designs a self-modifier to delete misclicked items and insert missed items, aiming to improve the recommendation accuracy, which is the strongest baseline in this paper.
    \end{itemize}
    
    \item Denoising sequential recommendation models:
    \begin{itemize}
        \item FMLP-Rec \cite{50-zhou} combines fast Fourier transform with an all-MLP architecture for denoising of sequential recommendation.
    \end{itemize}
    
\end{itemize}

For each baseline, we report the results on the test set according to the experimental results on the validation set. 
For STEAM and CLOUD, we report their experimental results on the modified sequences by default.

\subsection{Evaluation metrics and implementation}
We adopt two widely used TOP-K metrics to evaluate the performance of all the above sequential recommendation models: HR@K (hit ratio) and MRR@K (mean reciprocal rank) \cite{5-fang}, where the length of the recommendation list $K$ is set to 5, 10 or 20. The higher the HR and MRR, the more accurate the recommendation. 
To evaluate the privacy-preserving ability of CLOUD, we report the proportions of the three modification operations and the mean Jaccard similarity between the modified sequences and the raw sequences.

For all models, the experimental setup was jointly determined by the original article and the parameter tuning process. We use Xavier method \cite{66-lever} to initialize the model parameters and adopt Adam optimizer \cite{64-Kingma} to train the model, learning rate is set to 0.001 and embedding size is set to 64. We also apply gradient clipping \cite{65-Pascanu} with range $\left[-5, 5\right]$ during training. Like BERT4Rec \cite{25-sun}, we randomly draw 99 un-interacted items as negative samples for each validation and test item to train the recommender. In fact, we also found that the number of negative samples could affect the performance of recommender, which will be discussed in Section 5.4. 

For CLOUD, we set the number of heads in transformer to 1, the number of layers in the network to 1, and the dropout rate to 0.5. During the training phase, we randomly modify the raw sequence with probabilities of keep: $p_k=0.4$, deletion: $p_d=0.5$, insertion: $p_i=0.1$, and the mask probability $p_m=0.5$. We limit the maximum length of the raw sequence to 50, and if the length exceeds 50, the latest 50 records are kept. Up to 5 items can be inserted consecutively at each position of the sequence, and the maximum length of the modified sequence is limited to 60.

\section{Experimental results}
In this section, we conduct four sets of experiments to answer the research questions in Section 4.1.

\subsection{Privacy evaluation}
The primary goal of CLOUD is to reduce the leakage risk of real data by modifying the raw interaction sequence, so as to enhance the privacy-preserving of sequential recommendation. To answer RQ1, we count the proportion of three operations performed by CLOUD on all items in the test set, as well as the average Jaccard similarity between the modified sequences and the raw sequences. In addition, to explore the impact of each component in the item-wise modifier on the modification ratio, we also designed two variant models for CLOUD:
\begin{itemize}
    \item Variant-1 removes the copy mechanism in the item-wise modifier, and similar sequences are only used to calculate the probability distribution of performing three operations.
    \item Variant-2 calculates the probability distribution of performing three operations using only raw sequence, but retains the copy mechanism in the item-wise modifier.
\end{itemize}

\begin{table}[t]\footnotesize%
	\centering 
	\caption{Privacy evaluation of CLOUD and STEAM. Similarity denotes the average Jaccard similarity between the modified sequences and the raw sequences. Keep, Delete and Insert are the percentages of diferent types of modification operations.}
	\label{tab3}
    \renewcommand\tabcolsep{3pt}
    \renewcommand\arraystretch{1.5}
	\begin{tabular}{l cccc cccc cccc}
		\toprule
        & \multicolumn{4}{c}{Beauty}  & \multicolumn{4}{c}{Yelp}   & \multicolumn{4}{c}{Sports} \\
        \cmidrule(r){2-5}
        \cmidrule(r){6-9}
        \cmidrule(r){10-13}
		Model   &Similarity&Keep&Delete&Insert&Similarity&Keep&Delete&Insert&Similarity&Keep&Delete&Insert\\ 
		\midrule 
		STEAM & 0.7990 & 0.8825 & 0.0374 & 0.0801 & 0.8793 & 0.9951	& 0.0021 & 0.0028 &0.7856 & 0.9684 &	0.0261 & 0.0055\\
		Variant-1 & 0.6642 & 0.8088 & 0.0506 & 0.1406 & 0.6682 & 0.7761 & 0.0034 & 0.2205 & 0.6564 & 0.8380	& 0.0548 & 0.1072\\  
		Variant-2  & 0.6009 & 0.7528 & 0.0376 & 0.2096 & 0.5181	& 0.5739 & 0.0010 & 0.4251 & 0.6333 & 0.7725	& 0.0233 & 0.2042\\
		CLOUD & 0.5968 & 0.6880 & 0.0615 & 0.2505 & 0.4178 & 0.3343 & 0.0011 & 0.6646 & 0.6301 & 0.6975 & 0.0448 & 0.2577\\
		\bottomrule 
	\end{tabular}
\end{table}

As shown in Table \ref{tab3}, the experimental results are recorded from the epoch when the comprehensive performance of each model is optimized. Observing the experimental results, we can find: First, STEAM has a very low modification ratio, modifying only few items on the three datasets. In particular, the proportion of keep operations on Yelp is 99.51\%, and the modified sequence also has the highest Jaccard similarity with the raw sequence, which makes the privacy-preserving ability of STEAM very limited. Then, after calculating the probability distribution of operations by the shared representation of similar sequences and raw sequence, the item-wise modifier becomes more active, and the modification ratio of Variant-1 increases substantially compared to STEAM. Subsequently, the modification ratio of Variant-2 further increases, and the similarity continues to decline. This shows that the copy mechanism plays a relatively large role, which effectively encourages the item-wise modifier to modify the raw sequence more frequently.

Finally, CLOUD that combines these two parts has the strongest ability to modify sequences. Specifically, CLOUD modified more than 30\% of Beauty and Sports, and the average similarity of the sequences before and after modification was 59.68\% and 63.01\%, respectively. In particular, the modification ratio on Yelp is 66.57\%, and the Jaccard similarity is only 41.78\%, proving that CLOUD's collaborative-modifier is more powerful than STEAM's self-modifier. Further analysis shows that the change in modification ratio is mainly due to more frequent insertion operations, which is in line with our expectation. The purpose of introducing similar sequences is to hope that the item-wise modifier can learn the ability to perform insertion operations more actively with reference to similar sequences. In addition, from the perspective of privacy-preserving, insertion is more valuable than deletion. Because after performing the delete operation, the remaining interactions are still completely authentic. On the contrary, the insertion operation significantly enhances the confusion of the item sequence, making it difficult for the attacker to identify the real interaction record. Thus, it reduces the risk of attackers obtaining user related information at the source of the recommendation process.

\subsection{Recommendation evaluation}
Another goal of CLOUD is to enhance the privacy-preserving while maintaining superior recommendation performance. To answer RQ2, we conduct a set of experiments on three real datasets to compare the overall performance of CLOUD and the other eight baselines. The length of the recommendation list is set to 5,10,20, and the experimental results are shown in Table \ref{tab4}. By analyzing the experimental results, the following findings can be obtained.

\begin{table}[t]\footnotesize%
	\centering 
	\caption{Performance comparison of different methods on the real test sets. The best performance and the second best performance are denoted in bold and underlined fonts respectively.}
	\label{tab4}
    \renewcommand\tabcolsep{3pt}
    \renewcommand\arraystretch{1.1}
	\begin{tabular}{c| cccc cccc}
		\toprule
		Dataset & Model &HR@5&HR@10&HR@20&MRR@5&MRR@10&MRR@20&Sum\\ 
		\midrule 
        \multirow{10}{*}{Beauty} & 
        GRU4Rec & 0.3295 & 0.4259 & 0.5517 & 0.2163 & 0.2290 & 0.2376 & 1.9902 \\
        &BERT4Rec & 0.3666 & 0.4728 & 0.6011 & 0.2337 & 0.2478 & 0.2566 & 2.1790 \\
        &SASRec & 0.3657 & 0.4556 & 0.5763 & 0.2543 & 0.2662 & 0.2744 & 2.1928 \\
        &SRGNN & 0.3733 & 0.4765 & 0.5981 & 0.2514 & 0.2651 & 0.2735 & 2.2381 \\
        &FMLP-Rec & 0.3969 & 0.4872 & 0.5994 & 0.2800 & 0.2920 & 0.2997 & 2.3554 \\
        &CL4SRec & 0.4067 & 0.5056 & 0.6199 & 0.2785 & 0.2916 & 0.2994 & 2.4019 \\
        &DuoRec & 0.4094 & 0.5078 & 0.6200 & 0.2884 & 0.3015 & 0.3092 & 2.4366 \\
        &STEAM & \textbf{0.4284} & \underline{0.5256} & \textbf{0.6486} & \underline{0.2899} & \underline{0.3028} & \underline{0.3112} & \textbf{2.5069} \\
        &CLOUD & \underline{0.4252} & \textbf{0.5258} & \underline{0.6449} & \textbf{0.2910} & \textbf{0.3044} & \textbf{0.3125} & \underline{2.5041} \\
        &CLOUD-R & 0.4154& 0.5210& 0.6449 &0.2749& 0.2889& 0.2975& 2.4429 \\
        \midrule
        \multirow{10}{*}{Yelp} & 
        GRU4Rec & 0.5540 & 0.7657 & 0.9238 & 0.3222 & 0.3505 & 0.3619 & 3.2783 \\
        &BERT4Rec & 0.6118 & 0.7972 & 0.9235 & 0.3764 & 0.4013 & 0.4104 & 3.5208 \\
        &SASRec & 0.5824 & 0.7795 & 0.9206 & 0.3506 & 0.3771 & 0.3873 & 3.3978 \\
        &SRGNN & 0.5985 & 0.7895 & 0.9212 & 0.3674 & 0.3930 & 0.4026 & 3.4724 \\
        &FMLP-Rec & 0.6184 & 0.8075 & 0.9411 & 0.3838 & 0.4092 & 0.4188 & 3.5790 \\
        &CL4SRec & 0.6292 & 0.8223 & 0.9486 & 0.3991 & 0.4251 & 0.4343 & 3.6589 \\
        &DuoRec & 0.6400 & 0.8263 & \textbf{0.9564} & 0.4085 & 0.4334 & 0.4429 & 3.7078 \\
        &STEAM & \textbf{0.6683} & \textbf{0.8460} & 0.9511 & \underline{0.4293} & \underline{0.4531} & \underline{0.4607} & \textbf{3.8087} \\
        &CLOUD & 0.6653 & \underline{0.8414} & \underline{0.9525} & 0.4270 & 0.4506 & 0.4588 & 3.7958 \\
        &CLOUD-R & \underline{0.6674} & 0.8407 & 0.9511 & \textbf{0.4305} & \textbf{0.4538} & \textbf{0.4619} & \underline{3.8056} \\
        \midrule
        \multirow{10}{*}{Sports} & 
        GRU4Rec & 0.3057 & 0.4284 & 0.5801 & 0.1835 & 0.1997 & 0.2101 & 1.9077 \\
        &BERT4Rec & 0.3515 & 0.4791 & 0.6280 & 0.2154 & 0.2323 & 0.2426 & 2.1492 \\
        &SASRec & 0.3451 & 0.4619 & 0.6010 & 0.2190 & 0.2345 & 0.2441 & 2.1059 \\
        &SRGNN & 0.3592 & 0.4831 & 0.6303 & 0.2244 & 0.2408 & 0.2509 & 2.1889 \\
        &FMLP-Rec & 0.3766 & 0.4932 & 0.6318 & 0.2466 & 0.2620 & 0.2716 & 2.2820 \\
        &CL4SRec & 0.3935 & 0.5152 & 0.6562 & 0.2600 & 0.2762 & 0.2859 & 2.3872 \\
        &DuoRec & 0.3980 & 0.5192 & 0.6583 & 0.2597 & 0.2758 & 0.2854 & 2.3965 \\
        &STEAM & \textbf{0.4190} & \underline{0.5496} & \textbf{0.6916} & \textbf{0.2685} & \textbf{0.2858} & \textbf{0.2956} & \textbf{2.5103} \\
        &CLOUD & \underline{0.4186} & \textbf{0.5506} & 0.6881 & \underline{0.2678} & \underline{0.2854} & \underline{0.2949} & \underline{2.5056} \\
        &CLOUD-R & 0.4156 & 0.5474 & \underline{0.6884} & 0.2640 & 0.2816 & 0.2913 & 2.4886 \\
		\bottomrule 
	\end{tabular}
\end{table}

First, CLOUD still achieves state-of-the-art recommendation performance. In general, the recommendation accuracy of CLOUD is comparable to that of STEAM, and although it achieves the second experimental result in some metrics, it is only slightly different from STEAM (less than 0.5\%). Compared with other sequential recommendation algorithms, CLOUD has obvious performance advantages. Specifically, compared with the state-of-the-art SSL-based model DuoRec, CLOUD improves each metric by an average of 1.12 points on Beauty, 1.47 points on Yelp, and 1.82 points on Sports. In addition, we also report CLOUD-R as the recommendation results of CLOUD on the raw sequence. It can be found that CLOUD is slightly weaker than CLOUD-R on Yelp, but better on the other two datasets, which is in line with our expectations. This indicates that modifying the raw sequence can help the recommender to predict the next item more accurately, because the deletion and insertion correction tasks for the item-wise modifier can provide the recommender with powerful self-supervised signals to obtain better item representations and robust item correlations.

Second, observing the experimental results of other baselines, we can find: As the early Deep Learining-based sequential recommendation method, GRU4Rec has the lowest recommendation accuracy. BERT4Rec's performance on Yelp and Sports are significantly better than SASRec, which indicates that BERT4Rec's bidirectional transformer has more advantages in the experimental setting with a small number of negative samples. SRGNN adopts GNN to model sequential recommendation, and the overall performance is slightly better than BERT4Rec. In addition, SSL-based models CL4SRec and DuoRec outperform other vanilla models in all metrics, especially DuoRec is only weaker than STEAM, and even achieves the best performance of HR@20 on Yelp. This also shows that the self-supervised learning mechanism helps the model to obtain additional supervision signals from the sequence, and further improves the accuracy of recommendation.

Third, FMLP-Rec also achieves superior performance compared to BERT4Rec and SRGNN, even though it is a pure \acf{MLP} model rather than a more complex transformer or GNN architecture. This is because FMLP-Rec adopts \ac{FFTs} to filter out the noise information in the raw sequence, while the vanilla models may overfit on the noisy data due to their over-parameterized architectures \cite{66-lever,67-mehta}. CL4SRec and DuoRec are less affected by noise because self-supervised learning improves the robustness of the model. Finally, our study combines these two advantages to achieve better recommendation performance by deleting noise and inserting new items conducive to recommendation through self-supervised learning.

\begin{table}[t]\footnotesize%
	\centering 
	\caption{Robustness analysis. The table shows the performance comparison of different methods on the simulated test sets. The best performance is denoted in bold. * indicates that the performance of CLOUD against the best baseline is statistically significant based on a two-sided paired t-test with $p<0.01$.}
	\label{tab5}
    \renewcommand\tabcolsep{2pt}
    \renewcommand\arraystretch{1.1}
	\begin{tabular}{c|cccccccccc}
		\toprule
		Dataset & Model &HR@5&HR@10&HR@20&MRR@5&MRR@10&MRR@20&Sum & $Sum_{real}$ & dist\\ 
		\midrule 
        \multirow{5}{*}{Simulated Beauty}& 
        BERT4Rec & 0.3205 & 0.4230 & 0.5552 & 0.2022 & 0.2157 & 0.2247 & 1.9416 & 2.1790 & -10.89\% \\
        &FMLP-Rec & 0.3836 & 0.4722 & 0.5822 & 0.2743 & 0.2861 & 0.2936 & 2.2923 & 2.3554 & -2.68\% \\
        &CL4SRec & 0.3867 & 0.4809 & 0.5990 & 0.2778 & 0.2903 & 0.2984 & 2.3333 & 2.4019 & -2.85\% \\
        &DuoRec & 0.3961 & 0.4877 & 0.6025 & 0.2772 & 0.2894 & 0.2973 & 2.3504 & 2.4366 & -3.53\% \\
        &CLOUD & \textbf{0.4164*} & \textbf{0.5161*} & \textbf{0.6351*} & \textbf{0.2828*} & \textbf{0.2961*} & \textbf{0.3042*} & \textbf{2.4509} & \textbf{2.4429} & \textbf{0.32\%} \\
        \midrule 
        \multirow{5}{*}{Simulated Yelp}& 
        BERT4Rec & 0.5286 & 0.7130 & 0.8611 & 0.3129 & 0.3376 & 0.3482 & 3.1016 & 3.5208 & -11.90\% \\
        &FMLP-Rec & 0.5890 & 0.7890 & 0.9310 & 0.3619 & 0.3887 & 0.3990 & 3.4589 & 3.5790 & -3.35\% \\
        &CL4SRec & 0.6003 & 0.7910 & 0.9280 & 0.3721 & 0.3977 & 0.4077 & 3.4971 & 3.6589 & -4.42\% \\
        &DuoRec & 0.6132 & 0.8082 & \textbf{0.9518} & 0.3804 & 0.4064 & 0.4168 & 3.5769 & 3.7078 & -3.53\% \\
        &CLOUD & \textbf{0.6424*} & \textbf{0.8204*} & 0.9390 & \textbf{0.4059*} & \textbf{0.4299*} & \textbf{0.4385*} & 	\textbf{3.6763}	& \textbf{3.8056} & \textbf{-3.39\%} \\
        \midrule 
        \multirow{5}{*}{Simulated Sports}& 
        BERT4Rec & 0.3076 & 0.4332 & 0.5795 & 0.1843 & 0.2009 & 0.2110 & 1.9168 & 2.1492 & -10.81\% \\
        &FMLP-Rec & 0.3576 & 0.4736 & 0.6114 & 0.2328 & 0.2481 & 0.2576 & 2.1814 & 2.2820 & -4.40\% \\
        &CL4SRec & 0.3648 & 0.4854 & 0.6251 & 0.2404 & 0.2563 & 0.2659 & 2.2381 & 2.3872 & -6.24\% \\
        &DuoRec & 0.3747 & 0.4976 & 0.6360 & 0.2432 & 0.2594 & 0.2689 & 2.2802 & 2.3965 & -4.85\% \\
        &CLOUD & \textbf{0.4002*} & \textbf{0.5297*} & \textbf{0.6718*} & \textbf{0.2543*} & \textbf{0.2715*} & \textbf{0.2813*} & \textbf{2.4090} & \textbf{2.4886}	& \textbf{-3.19\%} \\
		\bottomrule 
	\end{tabular}
\end{table}

\subsection{Robustness analysis}
To answer RQ3, we next conduct a set of experiments on the simulated test sets to compare the recommendation performance of CLOUD with several powerful baselines. The experiments assume that there is an extreme case where user interaction records need to be randomly destroyed for privacy-preserving purpose. Based on this assumption, we perform keep, delete and insert operations on each item of the raw dataset with 40\%, 30\%, and 30\% probability to generate a new simulated dataset for testing the robustness of CLOUD. The experimental results are shown in Table \ref{tab5}, where $Sum$ denotes the total score of the evaluation metrics on the simulated test set and $Sum_{real}$ denotes the total score of the evaluation metrics on the real test set.

We can observe a similar performance ranking from Table \ref{tab5} as in Table \ref{tab4}. All models have noticeable performance degradation on the simulated test set due to the massive destruction of data. However, CLOUD still stably achieves the highest recommendation performance on the three datasets, and only lags behind DuoRec in the HR@20 on Yelp, which again proves the superiority of CLOUD’s performance.

In order to further analyze the robustness of each model, we also visually show the disturbance of their total score on the simulated test set compared with the total score on the real test set in Table \ref{tab5}, where $dist=(Sum-Sum_{real})/Sum_{real}$. Observing the experimental results, the following conclusions can be obtained: First, CLOUD is the most robust overall, achieving the lowest proportional decrease in total score on all three datasets. For CLOUD, $Sum_{real}$ is the metric score on the raw item sequences of the real test
set (see CLOUD-R in Table \ref{tab4}), because we hope to evaluate how CLOUD handles the simulated noises added into the real test set. Especially, the performance of CLOUD on the simulated Beauty is still better than CLOUD-R on the real Beauty. This indicates that benefiting from the powerful modification ability, CLOUD is able to make accurate recommendation even in the face of a lot of noisy data. Second, FMLP-Rec also shows strong competitiveness in robustness, achieving the second lowest performance degradation. As a denoising model, FMLP-Rec can effectively reduce the negative impact of randomly added noise in the simulated datasets, but its recommendation performance is significantly weaker than CLOUD. Finally, as a basic SR recommendation framework, BERT4Rec has the weakest performance and robustness on simulated datasets. Benefiting from the improvement of robustness by self-supervised learning, CL4SRec and DuoRec also show advanced performance on simulated datasets, but there is still a certain gap with CLOUD. 

In summary, corresponding to the case assumed at the beginning of this subsection, CLOUD is more robust than the above baselines in the face of random data destruction, which also means that CLOUD has greater potential to be applied to recommendation with privacy-preserving.

\subsection{Recommender analysis}
The core of CLOUD is the self-supervised modifier, and the recommender can be replaced flexibly. To answer RQ4, we finally conduct a set of experiments to observe the impact of different recommenders on the performance. Unlike the experiments conducted in RQ2, we evaluate performance on all negative samples (non-interacting items) instead of randomly sampling 99 items and design two variants for CLOUD:
\begin{itemize}
    \item CLOUD-uni uses a unidirectional transformer to train the model by masking the last position of sequences.
    \item CLOUD-bi uses a bidirectional transformer to train the model by masking arbitrary positions in the sequences with a probability of 0.5.
\end{itemize}

\begin{table}[t]\footnotesize%
	\centering 
	\caption{Evaluating the performance on all negative samples. The best performance is denoted in bold.}
	\label{tab6}
    \renewcommand\tabcolsep{3pt}
    \renewcommand\arraystretch{1.1}
	\begin{tabular}{c|c ccc ccc c}
		\toprule
		Dataset & Model &HR@5&HR@10&HR@20&MRR@5&MRR@10&MRR@20&Sum \\ 
		\midrule 
        \multirow{6}{*}{Beauty}& 
        BERT4Rec & 0.0351 & 0.0560 & 0.0859 & 0.0190 & 0.0217 & 0.0238 & 0.2418 \\
        &SASRec & 0.0543 & 0.0733 & 0.1021 & 0.0334 & 0.0359 & 0.0379 & 0.3372 \\
        &CL4SRec & 0.0630 & 0.0901 & 0.1254 & 0.0384 & 0.0419 & 0.0443 & 0.4035 \\
        &DuoRec & 0.0646 & 0.0909 & 0.1256 & 0.0405 & 0.0440 & 0.0463 & 0.4121 \\
        &CLOUD-uni & \textbf{0.0662} & \textbf{0.0931} & \textbf{0.1305} & \textbf{0.0407} & \textbf{0.0442} & \textbf{0.0468} & \textbf{0.4217} \\
        &CLOUD-bi & 0.0413 & 0.0661 & 0.1000 & 0.0218 & 0.025 & 0.0273 & 0.2817 \\
        \midrule 
        \multirow{6}{*}{Yelp}& 
        BERT4Rec & 0.0302 & 0.0488 & 0.0772 & 0.0168 & 0.0192 & 0.0211 & 0.2135 \\
        &SASRec & 0.0350 & 0.0487 & 0.0708 & 0.0229 & 0.0247 & 0.0262 & 0.2285 \\
        &CL4SRec & 0.0482 & 0.0724 & 0.1047 & 0.0296 & 0.0328 & 0.0350 & 0.3228 \\
        &DuoRec & \textbf{0.0485} & 0.0712 & 0.1041 & 0.0299 & 0.0329 & 0.0352 & 0.3221 \\
        &CLOUD-uni & \textbf{0.0485} & 0.0699 & 0.1033 & \textbf{0.0329} & \textbf{0.0357} & \textbf{0.0379} & \textbf{0.3285} \\
        &CLOUD-bi & 0.0478 & \textbf{0.0733} & \textbf{0.1111} & 0.0273 & 0.0306 & 0.0332 & 0.3235 \\
        \midrule 
        \multirow{6}{*}{Sports}& 
        BERT4Rec & 0.0132 & 0.0235 & 0.0403 & 0.0065 & 0.0078 & 0.0089 & 0.1004 \\
        &SASRec & 0.0262 & 0.0391 & 0.0574 & 0.0152 & 0.0169 & 0.0181 & 0.1731 \\
        &CL4SRec & 0.0346 & 0.0519 & 0.0760 & 0.0205 & 0.0227 & 0.0244 & 0.2304 \\
        &DuoRec & 0.0359 & 0.0518 & 0.0736 & 0.0214 & 0.0235 & 0.0250 & 0.2310 \\
        &CLOUD-uni & \textbf{0.0379} & \textbf{0.0547} & \textbf{0.0794} & \textbf{0.0225} & \textbf{0.0248} & \textbf{0.0264} & \textbf{0.2460} \\ 
        &CLOUD-bi & 0.0253 & 0.0408 & 0.0626 & 0.0132 & 0.0152 & 0.0167 & 0.1742 \\
		\bottomrule 
	\end{tabular}
\end{table}

Experimental results are shown in Table \ref{tab6}. Observing the experimental results, we can get different conclusion than RQ2. First, when evaluating the performance using all negative samples, BERT4Rec no longer has a performance advantage. Instead, it significantly lags behind SASRec on most metrics of the three datasets, and only achieves a lead on HR@10 and HR@20 on Yelp, which is the opposite of the performance ranking shown on 99 negative samples. Second, CL4SRec and DuoRec, which are also based on unidirectional transformers, achieve better performance than SASRec with the help of the self-supervised learning mechanism. Third, the recommendation performance of CLOUD-bi is significantly behind the other two SSL-based models, and even behind SASRec on Beauty. When we replace the recommender of CLOUD with unidirectional transformer, CLOUD-uni again shows the strongest recommendation performance in most metrics. This indicates that the recommender greatly affects the performance under different evaluation settings.

\begin{figure}[h]
  \centering
  \includegraphics[width=0.6\linewidth]{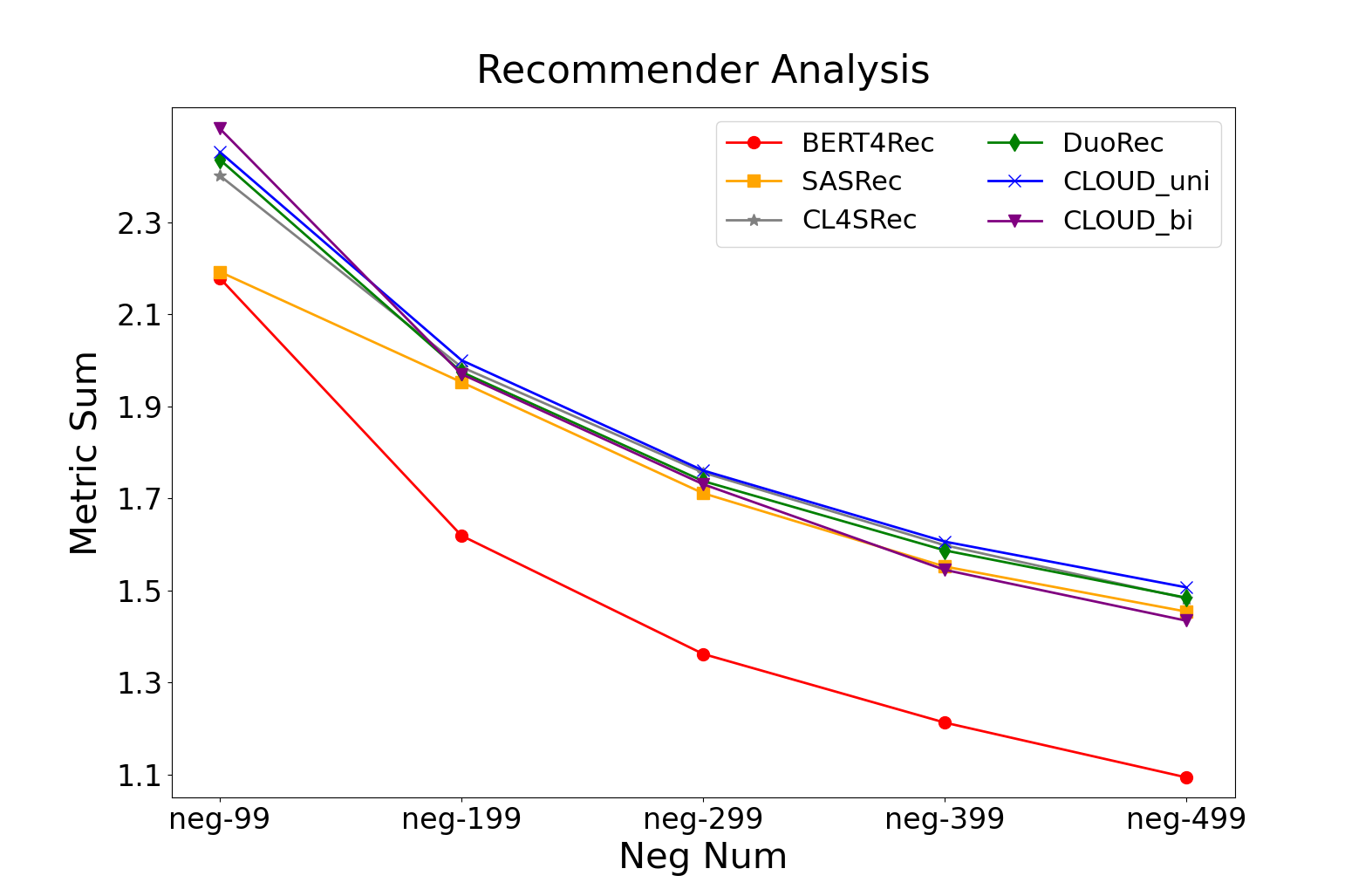}
  \caption{Recommender analysis with different negative samples on Beauty.}
  \Description{The figure is the result of RQ4}
  \label{fig6}
\end{figure}

In order to further observe this rule, we conduct an additional set of experiments on Beauty, setting the number of negative samples to 99, 199, 299, 399, 499, respectively, and observe the changing trend of the performance ranking of each model. The experimental results are shown in Figure \ref{fig6}. Here we only show the total scores of the six metrics. It can be found that as the number of negative samples increases, the performance disadvantage of Bert4Rec and CLOUD-bi becomes more obvious. When the number of negative samples grows to 499, CLOUD-bi is even weaker than SASRec. Meanwhile, unidirectional transformer-based CLOUD-uni achieves better performance ranking. Combined with the negative sample sampling strategy outlined in \cite{25-sun}, we believe that the bidirectional transformer based-recommender has more advantages in small-scale ranking, while the unidirectional transformer-based recommender is more suitable for large-scale pre-ranking. However, CLOUD always achieves the superior performance after selecting a suitable recommender.

\section{Discussion}
We would like to discuss why CLOUD can enhance the privacy-preserving of user data, and the limitations that still exist. 
At present, most centralized training-based recommendation models protect user privacy by adding noise to the representation or gradient, which makes it difficult for attackers to mine the users' relevant knowledge according to the recommendation output. 
In contrast, CLOUD can directly modify the users' interaction records before recommendation, and does not cause significant performance degradation. 
In practical applications, the trained item-wise modifier can be deployed locally, and the item-wise modifier only needs to deal with a small amount of historical interaction records. 
Thus, CLOUD does not put forward high requirements for the device and communication. 
The item-wise modifier and recommender only transmit the modified item sequences, which includes a large number of confused records, making it difficult for the attackers to distinguish users' real preference compared to using real interaction sequences.

However, CLOUD only enhances the privacy-preserving of sequential recommendation to a certain extent, it also has the following limitations: 
(1) The model still needs a part of real interaction records in the training phase. 
(2) Although the modification ratio is high, it still does not completely modify the raw interaction sequence. 
The first limitation is a common problem faced by centralized training-based recommendation models, as it is difficult to train an effective recommender without using real data. 
A common and effective practice is to train the model in a distributed manner, but this will cause a large burden on the user device and communication. 
In fact, according to the General Data Protection Regulation (GDPR)\footnote{\url{https://gdpr-info.eu/}}, the reasonable use of existing user data in marketing is acceptable \cite{39-han}. 
For the second limitation, we consider combining other existing privacy-preserving methods. 
For example, since CLOUD and differential privacy-based models implement privacy-preserving at different stages of the recommendation process, in theory, CLOUD can also add additional noise to the feature representation or gradient, depending on the practical needs to balance recommendation performance with privacy.

\section{Conclusion and future work}
In this work, we have proposed a new collaborative-confusion sequential recommender namely CLOUD to address the privacy-preserving requirements in sequential recommendation. 
We have designed a SSL-based item-wise modifier with copy mechanism, which can effectively modify the input interaction sequence, enhance the perplexity of the modified sequence, and improve the privacy protection of user interaction records. Extensive experiments on three real datasets have shown that CLOUD can achieve superior recommendation performance while protecting user privacy, which provides a new solution for the privacy-preserving research of sequential recommendation.

The main limitations of CLOUD are that it still requires real data during training, and the item-wise modifier does not modify all user interaction records. As to future work, we plan to add noise to item embeddings, combining CLOUD with differential privacy to achieve stronger privacy protection capabilities.
In addition, we also plan to adopt generative adversarial networks to construct a discriminator, and train a more deceptive modifier through the adversarial interaction between the modifier and the discriminator.

\begin{acks}
This research was supported by the China Scholarship Council, the National Key R\&D Program of China (grants No.2022YFC3303004, No.2020YFB1406704), the Natural Science Foundation of China (62102234, 62272274, 62202271, 61902219, 61972234, 62072279, 62322111, 62271289, 61971468), the Key Scientific and Technological Innovation Program of Shandong Province (2019JZZY010129), the Natural Science Foundation of Shandong Province (ZR2021QF129, ZR2021MF104, ZR2021MF113), the Natural Science Fund for Outstanding Young Scholars of Shandong Province (ZR2022YQ60), the Research Fund for the Taishan Scholar Project of Shandong Province (tsqn202306064), Innovation Ability Improvement Project of Small and Mediumsized Sci-tech Enterprises (No. 2021TSGC1084), and the Fundamental Research Funds of Shandong University.
All content represents the opinion of the authors, which is not necessarily shared or endorsed by their respective employers and/or sponsors.
\end{acks}

\bibliographystyle{ACM-Reference-Format}
\bibliography{CLOUD}

\end{document}